\documentclass[a4paper]{article}

\usepackage{INTERSPEECH2018}
\usepackage{ amssymb }
\usepackage{pifont}
\usepackage{subfigure}
\title{Analysis of Length Normalization in End-to-End Speaker Verification System}
\name{Weicheng Cai$^2$, Jinkun Chen$^2$, Ming Li$^{1}$\thanks{This research was funded in part by the National Natural Science Foundation of China (61401524,61773413), Natural Science Foundation of Guangzhou City (201707010363), Science and Technology Development Foundation of Guangdong Province (2017B090901045), National Key Research and Development Program (2016YFC0103905).}}
\address{
	 $^1$Data Science Research Center, Duke Kunshan University, Kunshan, China\\
	$^2$School of Electronics and Information Technology, Sun Yat-sen University, Guangzhou, China}
\email{ming.li369@dukekunshan.edu.cn}

\begin{document}

\maketitle
\begin{abstract}
The classical i-vectors and the latest end-to-end deep speaker embeddings are the two representative categories of utterance-level representations in automatic speaker verification systems. Traditionally, once i-vectors or deep speaker embeddings are extracted, we rely on an extra length normalization step to normalize the representations into unit-length hyperspace before back-end modeling. In this paper, we explore how the neural network learns length-normalized deep speaker embeddings in an end-to-end manner. To this end, we add a length normalization layer followed by a scale layer before the output layer of the common classification network. We conducted experiments on the verification task of the Voxceleb1 dataset. The results show that integrating this simple step in the end-to-end training pipeline significantly boosts the performance of speaker verification. In the testing stage of our $L_2$-normalized end-to-end system, a simple inner-product can achieve the state-of-the-art.
\end{abstract}

\section{Introduction}
Speaker recognition (SR) task can be defined as an utterance-level ``sequence-to-one" learning  problem. It is problem in that we are trying to retrieve information about an entire  utterance rather than specific word content~\cite{campbell2006support}. Moreover, there is  no constraint on the lexicon words thus training utterances  and testing segments may have completely different contents~\cite{Kinnunen2010An}. Therefore, given the input speech data, the goal may boil down to transform them into utterance-level representations, among them the inter-class variability is maximized and simultaneously the intra-class variability is minimized~\cite{hansen2015speaker}.

Typically, SR can be categorized as speaker identification (SID) task 
and speaker verification (SV) task~\cite{Reynolds1995Robust}. The former classifies a speaker to a specific identity, while the latter determines whether a pair of utterances belongs to the same person.   For the open-set
protocol, speaker identities in the testing set are usually disjoint from the ones in 
training set, which makes the SV  more challenging yet closer to
practice. Since it is impossible to classify testing utterances to known
identities in training set, we need to map speakers to a discriminative
feature space.  In this scenario, open-set SV is essentially a metric learning problem, where
the key  is to learn discriminative large-margin speaker embeddings.

There are generally two categories commonly used to obtain utterance-level speaker representations. The first consists of series of separated statistical models. The represent is the classical i-vector approach~\cite{dehak2010front}. Firstly, frame-level feature sequences are extracted from raw audio signals. Then, selected feature frames in training dataset are grouped together to estimate a Gaussian Mixture Model (GMM) based universal background model (UBM)~\cite{Reynolds2000Speaker}. Sufficient statistics of each utterance on the UBM is accumulated, and a factor analysis based i-vector extractor is trained to project the statistics into a low-rank total variability subspace~\cite{dehak2010front}.  

The other category relies on a model trained by a downstream procedure through end-to-end deep neural network~\cite{gonzalez2014automatic,1705.02304,Snyder2017Deep,Cai_2018_Odyssey}. First, in the same way as the i-vector approach, frame-level feature sequences are extracted as well. Then an automatic frame-level feature extractor such as convolution neural network (CNN)~\cite{1705.02304, caie2e_iccasp18},  time-delay neural network (TDNN)~\cite{Snyder2017Deep} or Long Short Term Memory (LSTM) network~\cite{gonzalez2014automatic, bredin2017tristounet} is designated to get high-level abstract representation. Afterward, a statistic pooling~\cite{Snyder2017Deep} or  encoding layer~\cite{cailde_iccasp18} is built on top to extract the fixed-dimensional utterance-level representation. This utterance-level representation can be further processed by fully-connected (FC) layer, and finally connected with an output layer. All the components in the end-to-end pipeline are jointly learned with a unified loss function.

In classical i-vector approach, an extra length normalization step is necessary to normalize the representations into unit-length hyperspace before back-end modeling~\cite{garcia2011analysis}. When it turns into end-to-end system, once we have extracted deep speaker embeddings from theneural network, such as x-vector~\cite{xvector},  this length normalization step is also required when calculating pairwise scores. 
\begin{figure}[b]
	\centering
	\includegraphics[width=\columnwidth]{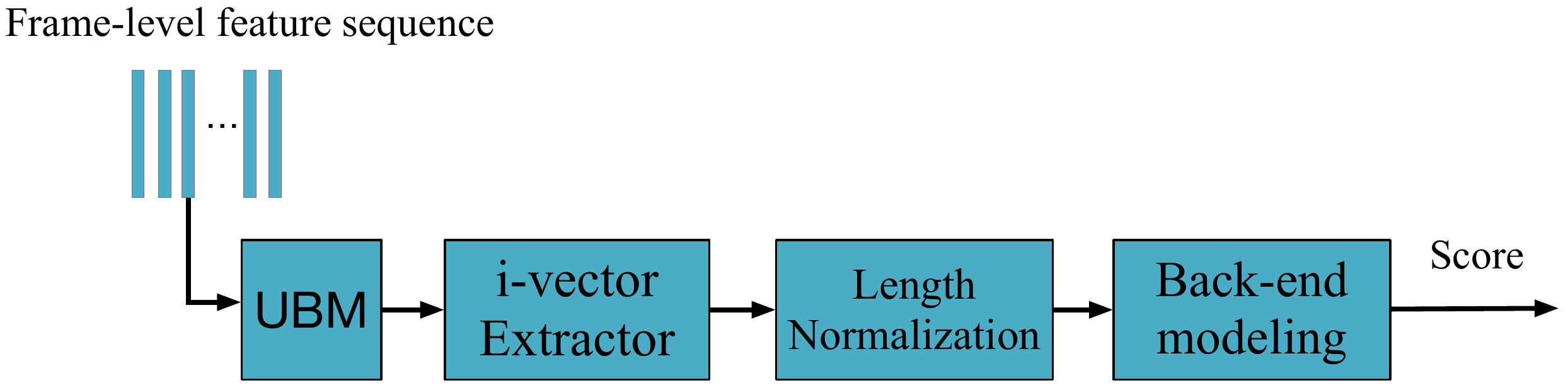}
	\caption{Length normalization in conventional i-vector system}\label{fig:ivector}
\end{figure}
\begin{figure*}[tb]
	\centering
	\includegraphics[width=0.8\textwidth]{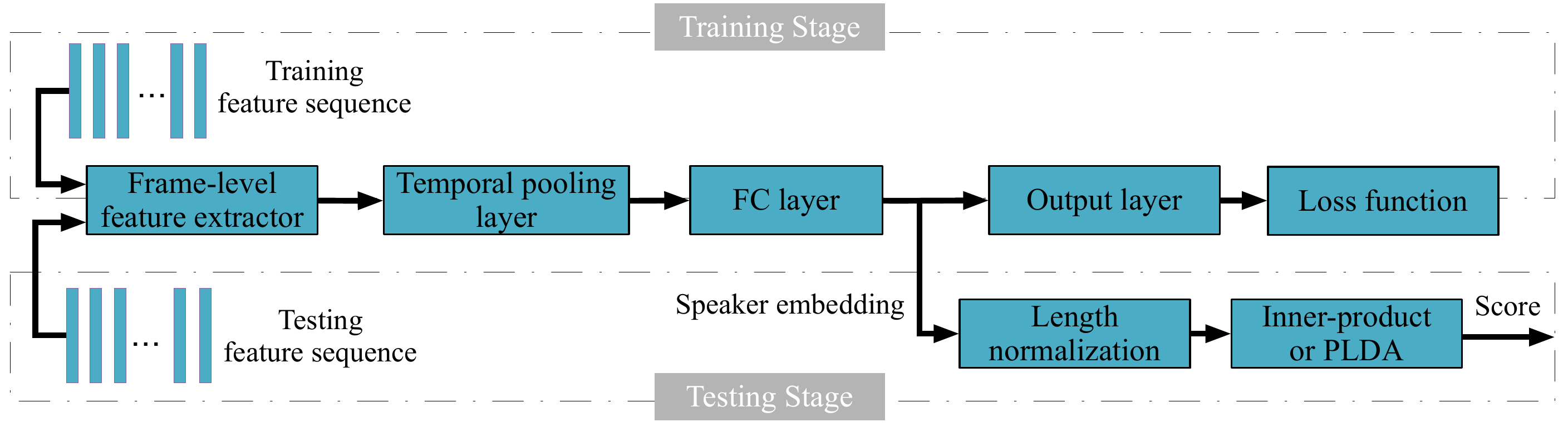}
	\caption{Length normalization in common deep speaker embedding system.}\label{fig:basic}
\end{figure*}

In this paper, we explore end-to-end SV system where  length normalization step is built-in inherently within the deep neural network. Therefore, the neural network can learn speaker embeddings being length-normalized in an end-to-end manner.

\section{Related works}

\subsection{Length normalization in i-vector approach}
\label{sec:ivector}

Length normalization has been analyzed and proved to be an effective strategy for SR, but limited in conventional i-vector approach~\cite{garcia2011analysis}. As demonstrated in Fig.~\ref{fig:ivector}, this simple non-linear transformation on i-vector has been the de facto standard before back-end modeling~\cite{bousquet2012variance,garcia2015insights}.  

For closed-set SID task,  length normalization followed by logistic regression or support vector machine is commonly adopted to get the posterior probabilities for the speaker categories. For open-set SV task, cosine similarity or length normalization followed by probabilistic
linear discriminant analysis (PLDA) scoring~\cite{prince2007probabilistic, kenny2010bayesian} modeling is widely used to get the final pairwise scores. The cosine similarity is a similarity measure which is independent of magnitude, it can be seen as the length-normalized version of inner-product of two vectors. In these above systems, front-end i-vector modeling, length normalization step, and back-end modeling are all independent of each other and performed separately.

\subsection{Length normalization in end-to-end system with triplet loss}
Some previous works in~\cite{1705.02304, bredin2017tristounet, Zhang2017} introduced triplet loss~\cite{Schroff2015FaceNet} and successfully trained models with the features being normalized in an end-to-end fashion. They all explicitly treat the open-set  SV task as a metric learning problem. This kind of triplet loss approach naturally requires length normalization step to compute the distance of normalized unit vectors. 

 However, a neural network trained with triplet loss requires carefully designed triplet mining procedure. This procedure is non-trivial,  both time-consuming and performance-sensitive~\cite{liu2017sphereface}.  Besides, many closed-set tasks like SID are equal to classification problem, it is intuitively not necessary to implement triplets mining procedure and explicitly treat them as metric learning problem. Therefore, we concentrate our attention on  general scenario with common classification network. This means the units in the output layer are equal to the pre-collected speaker categories in training set.

\subsection{Length normalization in  common end-to-end deep speaker embedding system}
For open-set SV task,  since it is impossible to classify testing utterances to known
identities in training set, the end-to-end classification network plays role as  an automatic speaker embedding extractor, as demonstrated in Fig.~\ref{fig:basic}.  Once deep speaker embeddings (e.g. x-vectors) are extracted, just the same as in i-vector approach, cosine similarity or length normalization followed by PLDA is commonly required to get the final pairwise scores. It's noticed that no matter in cosine similarity or PLDA modeling, the length normalization is an extra step performed on the extracted speaker embeddings, and out of end-to-end manner.

\section{Deep length normalization}
As described in section~\ref{sec:ivector}, back-end modeling in conventional i-vector approach usually performs on the unit-length hyperspace. When it turns into end-to-end deep neural network, however,  in practice the back-end softmax classifier commonly adopts the inner-product based FC layer without normalization. It means that if we want to perform cosine similarity or PLDA on the extracted deep speaker embeddings, such as the representative x-vectors, we should manually normalize them with unit-length first. 

It motivates us that whether it is possible to learn the deep speaker embeddings being length-normalized in an end-to-end manner within common classification network. One might wonder the substantial difference between length normalization in an end-to-end manner or out of end-to-end manner. This issue has been studied by~\cite{ranjan2017l2, wang2017normface} in computer vision community. The effect of deep length normalization is equivalent to adding an $L_2$-constraint on the original loss function. With deep speaker  embeddings being length-normalized inherently in an end-to-end manner,  our optimization object requires not only the speaker embeddings being separated, but also constrained on a small unit hyperspace.  This makes it more difficult to train the network, but in the other side, could greatly enhance its generalization capability.
\begin{figure*}[tb]
	\centering
	\includegraphics[width=0.88\textwidth]{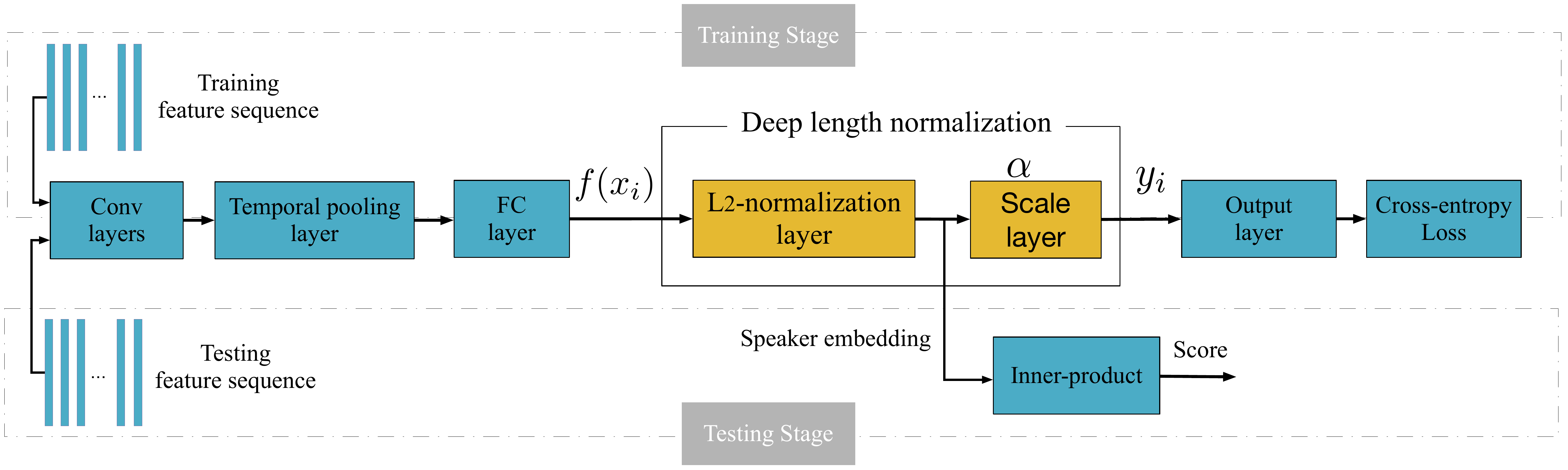}
	\caption{Deep length normalization in end-to-end speaker verification system: before the final output layer,  an $L_2$-normalization layer followed by a scale layer is added. Therefore, deep speaker embeddings can be inherently length-normalized in an end-to-end manner}\label{fig:l2norm}.
\end{figure*}

To this end, a naive practice is just to add an  $L_2$-normalization layer before the output layer. However, we find that the training process may not converge and lead to rather poor performance, especially when the number of output categories is very large. The reason might be that the surface area of the  unit-length hypersphere would have not enough room to not only accommodate so many speaker embeddings, but also allow each category of them to be separable. 

As done in~\cite{ranjan2017l2, wang2017normface},  we introduce a scale parameter $\alpha$ to shape the length-normalized speaker embeddings into suitable radius.  The scale layer can scale the unit-length speaker embeddings into a fixed radius given by the parameter $\alpha$. Therefore, the complete formula of our introduced deep length normalization can be expressed as 

\begin{equation}
\bm{y_i} = \alpha \times   \frac{f(\bm{x}_i)}{\left \| f(\bm{x}_i) \right \|_2}
\end{equation}
where $\bm{x_i}$ is the  $i^{\text {th}}$ input data sequence in the batch, $f(\bm{x_i})$ is the corresponding output of the penultimate layer of the network, and $\bm{v_i}$ is the deep normalized embedding. 

Our end-to-end system architecture with deep feature normalization is demonstrated in Fig.~\ref{fig:l2norm}. The length-normalized speaker embedding can be directly fed into the output layer, and all the components in the network are optimized jointly with a unified cross-entropy loss function:

\begin{equation}
\ell_{s}=-\frac{1}{M}\sum_{i=1}^M \log\frac{e^{\bm{W}_{c_i}^T\bm{y_i}+\bm{b}_{c_i}}}{\sum_{j=1}^{C}e^{\bm{W}_j^T\bm{y_j}+\bm{b}_j}} 
\end{equation}
where $M$ is the training batch size, $C$ is the output categories, $\bm{y_i}$ is the deep normalized embedding, $c_i$ is the corresponding ground truth label, and $\bm{W}$ and $\bm{b}$ are the weights and bias for the last layer of the network which acts as a back-end classifier.

In total,  only a single scalar parameter $\alpha$
is introduced, and it can be inherently trained with other components of the
network together. This scale parameter $\alpha$ has a crucial impact on the performance since it determines the radius of the length-normalized hyperspace. The network could have stronger $L_2$-constraint on the small radius hyperspace with smaller $\alpha$, but faces the risk of not convergent. 

Therefore,  it is vital to choose appropriate $\alpha$ and normalize the feature into hyperspace with suitable radius. For elegance, we may prefer to make the parameter $\alpha$ automatically learned by back-propagation.  However, because the cross-entropy loss function only takes into account whether it the  speaker embeddings are separated correctly, it is apt to increases the value of $\alpha$ to meet the demand. Therefore, the value of $\alpha$  learned by the network might always be high, which results in a relaxed $L_2$-constraint~\cite{ranjan2017l2}. 

A better practice considers $\alpha$ as a hype-parameter, and fix it with a low-value constant in order to enlarge the $L_2$-constraint. However, too small $\alpha$ for large number of categories may lead to the unconverged case. Hence, we should find an optimal balance point for $\alpha$.

Given the number of categories $C$ for a training dataset, in order to achieve a classification probability score of $p$, the authors in~\cite{ranjan2017l2} give the formulation of  theoretical lower bound on $\alpha$ by 
\begin{equation}
\alpha_{low} = \log \frac{p(C - 2)} {1 - p}
\end{equation}

At the testing stage, speaker embeddings are extracted after the $L_2$-normalization layer. Since the embeddings have already been normalized to unit length, a simple inner-product or PLDA can be adopted to get the final similarity scores. 

\section{Experiments}
\subsection{Data description}
Voxceleb1 is a large scale text-independent SR dataset collected ``in the wild", which contains over 100,000 utterances from 1251 celebrities~\cite{Nagrani17}. We focus on its open-set verification task.

There are totally 1211 celebrities in the development dataset. The testing dataset contains 4715 utterances from the rest 40 celebrities. There are totally 37720 pairs of trials including 18860 pairs of  true trials. To evaluate the system performance, we report results in terms of equal error-rate (EER) and the minimum of the normalized detection cost function (minDCF) at $P_{Target}$ = 0.01 and $P_{Target}$ = 0.001,  as shown in Table~\ref{table:voxceleb_verify} and Table~\ref{table:alpha}.

\subsection{Referenced i-vector system}
We build a referenced  i-vector system based on the Kaldi toolkit~\cite{Povey_ASRU2011}. Firstly, 20-dimensional  mel-frequency cepstral coefficients (MFCC) is augmented with their delta and double delta coefficients, making 60-dimensional MFCC feature vectors. Then, a frame-level energy-based voice activity detection (VAD) selects features corresponding to speech frames.  A 2048-components full covariance GMM UBM is trained, along with a 400-dimensional i-vector extractor and full rank PLDA.

\begin{table}[tb]
	\normalsize
	\centering
	\caption{ Baseline end-to-end system architecture}
	\label{resnetconfig}
	\resizebox{0.98\columnwidth}{!}{
		\renewcommand\arraystretch{1.3}
		\begin{tabular}{|c|c|c|c|c|}
			\hline
			Layer               & Output size            & Downsample      &  Channels     &  Blocks      \\ \hline
			Conv1                     & 64 $\times$ $L$                & False& 16       & -    \\ \hline
			Res1                & 64$\times$  $L$               & False& 16& 3    \\ \hline
			Res2                & 32 $\times$ $\frac{L}{2}$               & True& 32& 4    \\ \hline
			Res3                & 16 $\times$ $\frac{L}{4}$              & True& 64& 6   \\ \hline
			Res4                & 8 $\times$ $\frac{L}{8} $              & True& 128& 3    \\ \hline
			Average pool         & 128 &-&-&- \\ \hline		
			FC (embedding)         & 128 &-&-&- \\ \hline	
			Output         & speaker categories &-&-&- \\ \hline	
	\end{tabular}}
\end{table}

\begin{table*} [tb] 
	\caption{  Voxceleb1 open-set verification task performance,  in comparing the  effect of our introduced deep length normalization strategy and traditional extra length normalization step (lower is better)}
	\centerline {
		\resizebox{\textwidth}{20mm}{
			\begin{tabular}{c c c c  c c c c c} 	
				\hline
				System Description& Deep $L_2$-norm &   Traditional $L_2$-norm &  Similarity Metric&minDCF$10^{-2}$&minDCF$10^{-3}$&$EER(\%)$\\
				\hline
				i-vector +  inner-product   & N/A& \ding{55} &inner-product&0.736&0.800&13.80\\
				i-vector + cosine   & N/A& \ding{51}  &inner-product&0.681&0.771&13.80\\
				i-vector + PLDA  &N/A&\ding{55}&PLDA&0.488&0.639&5.48\\	
				i-vector + $L_2$-norm + PLDA&N/A  &\ding{51} &PLDA&0.484&0.627&5.41\\
				\hline		
				Deep embedding + inner-product   &\ding{55}&\ding{55}&inner-product&0.758&0.888&7.42\\
				Deep embedding+ cosine   &\ding{55}&\ding{51} &inner-product&0.553&0.713&5.48\\					
				Deep embedding+ PLDA  &\ding{55}& \ding{55}  &PLDA&0.524&0.739&5.90\\  
				Deep embedding + $L_2$-norm + PLDA  &\ding{55}&\ding{51} &PLDA&0.545&0.733&5.21\\
				\hline
				\textbf{$\bm{L_2}$-normalized deep embedding + inner-product} &\ding{51} &\ding{55}&inner-product&\textbf{0.475}&\textbf{0.586}&5.01\\ 
				$L_2$-normalized deep embedding + PLDA &\ding{51}&\ding{55}&PLDA&0.525&0.694&\textbf{4.74}\\ 
				\hline
	\end{tabular}}}
	\label{table:voxceleb_verify}
\end{table*}

\subsection{End-to-end system}
Audio is converted to 64-dimensional log mel-filterbank energies with a frame-length of 25 ms, mean-normalized over a sliding window of up to 3 seconds. A frame-level energy-based voice activity detection (VAD) selects features corresponding to speech frames.  In order to get higher level abstract representation, we design a deep convolutional neural network (CNN) based on the well-known ResNet-34 architecture~\cite{He2016Deep}, as described in Table~\ref{resnetconfig}. Followed by the front-end deep CNN, we adopt the simplest average pooling layer to extract the utterance-level mean statistics. Therefore, given input data sequence of shape $64\times L$,  where $L$ denotes variable-length data frames, we finally get 128-dimensional utterance-level representation.

The model is trained with a mini-batch size of 128,  using typical stochastic gradient descent with momentum 0.9 and weight decay 1e-4.  The learning rate is set to 0.1, 0.01, 0.001 and is switched when the training loss plateaus.  For each training step,  an integer $L$ within $\left[ 300  \textrm{,}  800 \right]$  interval is randomly generated, and each data in the mini-batch is cropped or extended to $L$ frames. After model training finished, the 128-dimensional speaker embeddings are extracted after the penultimate layer of neural network.

\begin{table} [tb] 
	\caption{    Verification performance on VoxCeleb1 for various scale parameter $\alpha$ (lower is better) }
	\centerline {
			\resizebox{\columnwidth}{21mm}{
		\begin{tabular}{c c c c} 
			\hline
			System Description&minDCF$10^{-2}$&minDCF$10^{-3}$&$EER(\%)$\\
			\hline	
			Deep embedding baseline&0.553&0.713&5.48\\
			\hline	
			fixed $\alpha=1 $ & 0.922&0.967&10.18\\
			fixed $\alpha=4$ &0.601&0.828&6.36\\
			fixed $\alpha=8$ &0.515&0.687&5.49\\
			\textbf{	fixed $\bm{\alpha=12}$ }&\textbf{0.475}&\textbf{0.586}&\textbf{5.01}\\
			fixed $\alpha=16$ &0.499 &0.596&5.32\\
			fixed $\alpha=20$& 0.503 &0.637&5.46\\
			fixed $\alpha=24$&0.502 &0.638& 5.54\\
			fixed $\alpha=28$&0.497&0.640&5.52\\
			trained $\alpha=26.1$&0.486&0.599&5.60\\
			\hline
	\end{tabular}}}
	\label{table:alpha}
\end{table}
\begin{figure}[tb]
	\centering
	\subfigure[Performance in terms of minDCF$10^{-2}$]{
		\label{fig:subfig:a} 
		\includegraphics[width=0.33\textwidth]{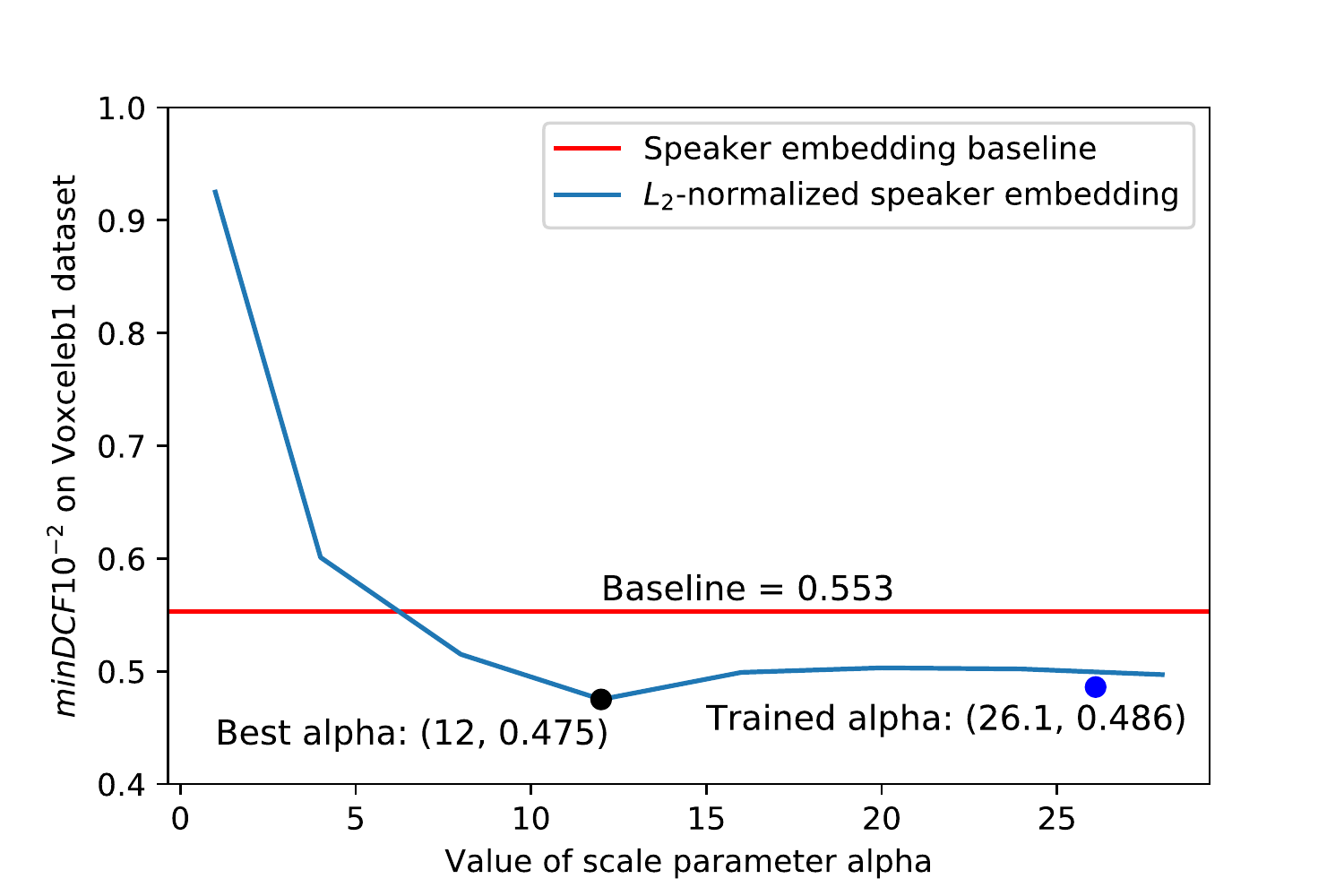}}
	\hspace{0.01in}
	\subfigure[Performance in terms of minDCF$10^{-3}$]{
		\label{fig:subfig:b} 
		\includegraphics[width=0.33\textwidth]{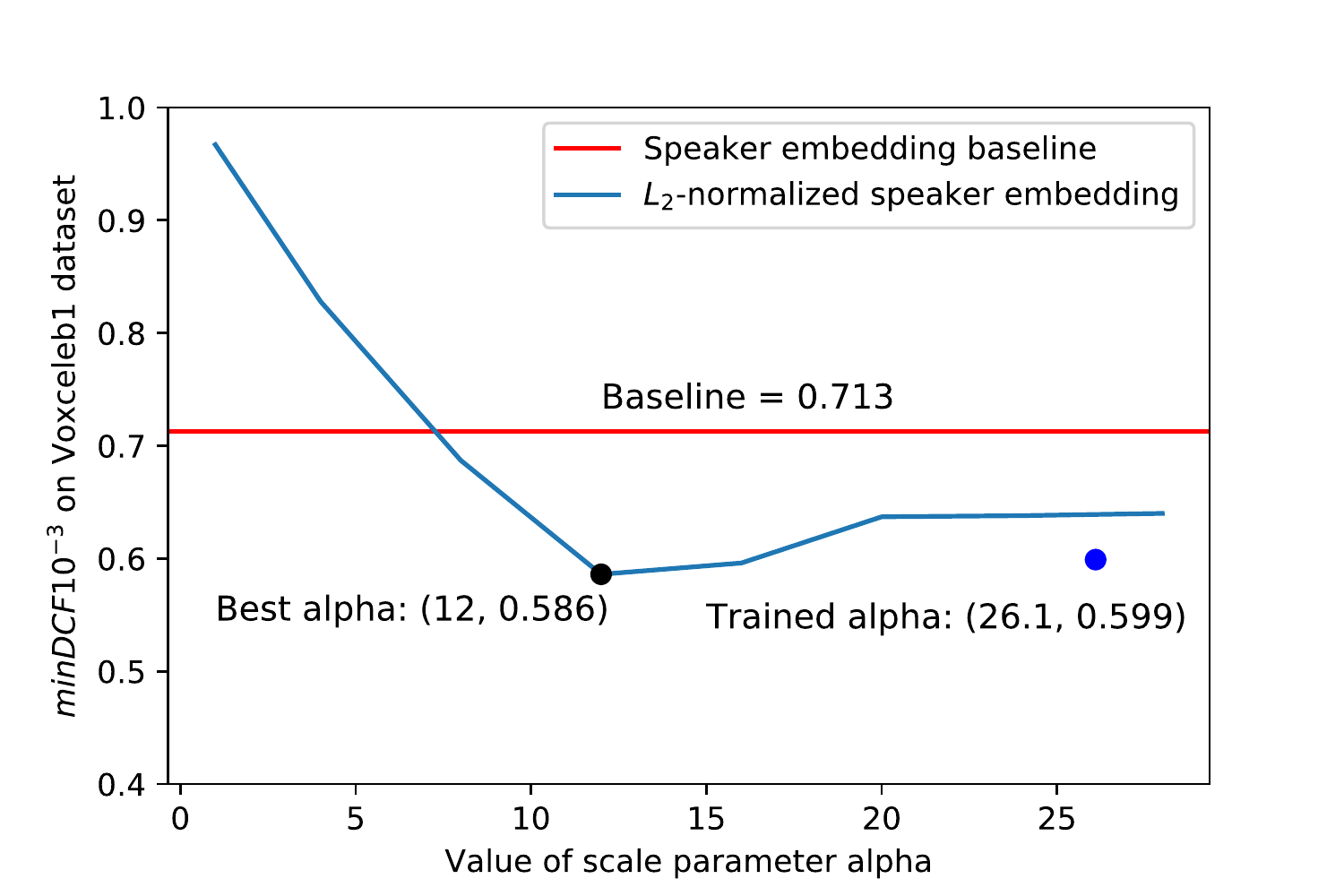}}
	\caption{ Performance tendency curve considering various $\alpha$ }\label{fig:performance}
\end{figure}

\subsection{Evaluation}

We first investigate the setting of scale parameter $\alpha$.  For those systems in Table~\ref{table:alpha} and Fig.~\ref{fig:performance},  the cosine similarity or equivalently $L_2$-normalized
inner-product is adopted to measure the similarities between
speaker embeddings. From Fig.~\ref{fig:performance}, we can observe the proposed $L_2$-normalized deep embedding system achieves the best minDCF of 0.475, 0.586 and EER of 5.01\%, which outperforms the baseline system significantly. According to Equation (3),  for speaker categories $C$ of 1211 and probability score $p$ of 0.9, the theoretical lower bound of $\alpha$ is 9. The performance is poor when $\alpha$ is below the lower bound and stable with $\alpha$ higher than the lower bound. The best $\alpha$ in our experiment is 12, which is slightly larger than the lower bound.

We further compare the effect of deep length normalization strategy and traditional extra length normalization in the whole SV pipeline. The results are shown in Table~\ref{table:voxceleb_verify}. No matter in i-vector or baseline deep speaker embedding systems, extra length normalization step followed by PLDA scoring achieves the best performance.  When it turns into $L_2$-normalized deep speaker embedding systems, since the speaker embeddings extracted from the neural network have already been normalized to unit length, we need no more extra length normalization step. In the testing stage, a simple inner-product achieves the best performance,  even slightly better than the PLDA scoring result. It might be the reason that our $L_2$-normalized speaker embeddings are highly optimized, which could be incompatible with the objective function introduced by  PLDA.


\section{Conclusions}
\label{sec:conclusion}

In this paper, we explore a deep length normalization strategy in end-to-end SV system. We add an $L_2$-normalization layer followed by a scale layer before the output layer of the deep neural network. This simple yet efficient strategy makes the learned deep speaker embeddings being normalized in an end-to-end manner. The value of scale parameter $\alpha$ is crucial to the system performance especially when the number of output categories is large. Experiments show that system performance could be significantly improved by setting a proper value of $\alpha$.  In the testing stage of an $L_2$-normalized deep embedding system, a simple inner-product can achieve the state-of-the-art.

\newpage
\bibliographystyle{IEEEtran}
\bibliography{l2norm_sv}


\end{document}